\documentclass[pra,twocolumn,amssymb]{revtex4}
\usepackage{graphicx,color,epsfig}
\begin{document}

\title{HITRAP: A facility at GSI for highly charged ions \footnote{dedicated to Ingvar Lindgren on the occasion of his 75$^{th}$ birthday}}

\author{H.-J. Kluge$^{a,b}$}
\email{J.Kluge@gsi.de} \homepage{www.gsi.de}
\author{T. Beier$^a$}
\author{K. Blaum$^{a,c}$}
\author{L. Dahl$^a$}
\author{S. Eliseev$^a$}
\author{F. Herfurth$^a$}
\author{B. Hofmann$^a$}
\author{O. Kester$^a$}
\author{S. Koszudowski$^a$}
\author{C. Kozhuharov$^a$}
\author{G. Maero$^a$}
\author{W. N\"ortersh\"auser$^{a,c}$}
\author{J. Pfister$^{a,d}$}
\author{W. Quint$^{a}$}
\author{U. Ratzinger$^d$}
\author{A. Schempp$^d$}
\author{R. Schuch$^e$}
\author{T. St\"ohlker$^a$}
\author{R.C. Thompson$^f$}
\author{M. Vogel$^a$}
\author{G. Vorobjev$^a$}
\author{D.F.A. Winters$^a$}
\author{G. Werth$^c$}

\affiliation{$^a$Gesellschaft f\"ur Schwerionenforschung, D-64291
Darmstadt, Germany} \affiliation{$^b$Universit\"at Heidelberg,
Physikalisches Institut, D-69120 Heidelberg, Germany}
\affiliation{$^c$Johannes Gutenberg-Universit\"at Mainz, D-55099
Mainz, Germany} \affiliation{$^d$Johann Wolfgang
Goethe-Universit\"at Frankfurt, D-60438 Frankfurt, Germany}
\affiliation{$^e$Department of Atomic Physics, Frescativ\"agen 24,
Stockholm University, Sweden} \affiliation{$^f$Imperial College
London, SW7 2AZ London, United Kingdom}
\date{\today}

\begin{abstract}
An overview and status report of the new trapping facility for
highly charged ions at the Gesellschaft f\"ur Schwerionenforschung
is presented. The construction of this facility started in 2005
and is expected to be completed in 2008. Once operational, highly
charged ions will be loaded from the experimental storage ring ESR
into the HITRAP facility, where they are decelerated and cooled.
The kinetic energy of the initially fast ions is reduced by more
than fourteen orders of magnitude and their thermal energy is
cooled to cryogenic temperatures. The cold ions are then delivered
to a broad range of atomic physics experiments.
\end{abstract}

\maketitle
\tableofcontents

\begin{figure}[!tb]
\centering
\includegraphics[width=8.5cm]{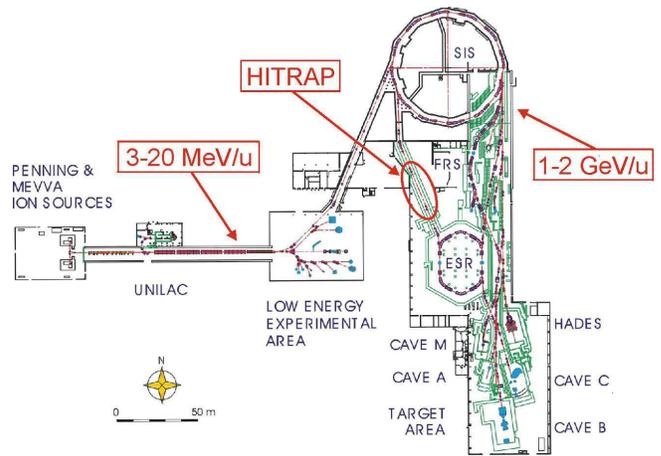}
\caption{The present GSI facility: highly charged ions are
produced by a variety of different ion sources and accelerated by
the linear accelerator (UNILAC) to several MeV/u. These fast ions
can be used for low-energy experiments or can be injected into the
heavy-ion synchrotron (SIS) where they are further accelerated.
The SIS feeds the fragment separator (FRS), the experimental
storage ring (ESR), or the fixed-target experiments in the
different caves including that for heavy-ion tumor therapy. The
HITRAP facility will be inserted into the re-injection channel,
which can be used to feed the SIS with cooled highly charged ions
from the ESR.} \label{fig1}
\end{figure}

\section{The present GSI facility}
The UNILAC at the Gesellschaft f\"ur Schwerionenforschung (GSI)
produced its first heavy-ion beam in 1975. Today, the current GSI
facility can provide quasi-continuous or pulsed beams of ions with
practically any charge state, ranging up to U$^{92+}$, with
kinetic energies of several MeV or GeV per nucleon.
Figure~\ref{fig1} shows schematically the present GSI accelerator
facility. The ions are produced by Penning, electron cyclotron
resonance (ECR) and metal vapour vacuum arc (MEVVA) ion sources,
and are then accelerated by a linear accelerator (UNILAC), which
is roughly 120 meters long, to the first range of kinetic
energies, {\it i.e.} up to several MeV/u. After the first
acceleration stage, the ions can be delivered to a series of
low-energy experiments. One example is SHIPTRAP, its name
originating from the SHIP velocity filter that was used to
discover six new elements: $_{107}^{264}$Bh (Bohrium, 1981),
$_{108}^{269}$Hs (Hassium, 1984), $_{109}^{268}$Mt (Meitnerium,
1982), $_{110}^{271}$Ds (Darmstadtium, 1994), $_{111}^{272}$Rg
(R\"ontgenium, 1994), and $_{112}^{277}$Uub (Ununbium, 1996).
SHIPTRAP is used to perform high-precision mass measurements of
trapped (unstable) radionuclides with a relative mass uncertainty
of $\delta m/m \approx 10^{-8}$.

In the heavy-ion synchrotron SIS the ions are accelerated to
kinetic energies of up to 2 GeV per nucleon for atomic and nuclear
physics experiments as well as heavy-ion tumor therapy. When using
highly charged ions, ultra-high vacuum (UHV) conditions are
crucial in order to avoid charge changing collisions. In the
UNILAC, where the ions' charge states are not yet very high,
pressures of around $10^{-6}$ mbar are sufficient. However, in the
heavy-ion synchrotron SIS (216 m circumference) and in the
experimental storage ring ESR (108 m circumference), for which the
charge states are increased by shooting the highly-energetic ions
through a stripper foil, the pressure is in the $10^{-11}$ mbar
regime.

\section{The HITRAP project}
\begin{figure}[!bt]
\centering
\includegraphics[width=8.5cm]{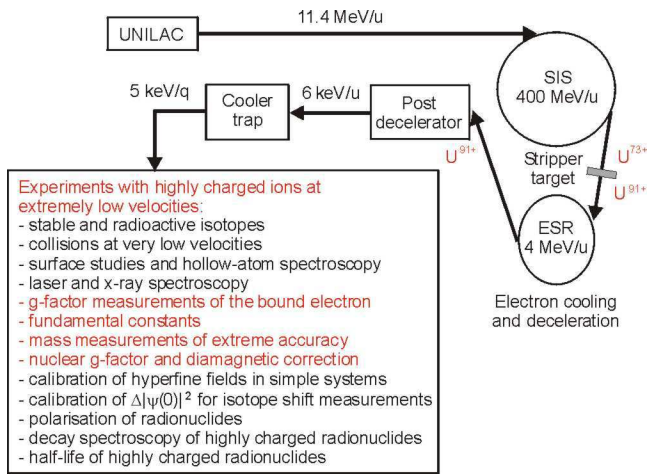}
\caption{Schematic of the HITRAP project: deceleration, trapping,
and cooling of highly charged ions for a large variety of atomic
physics experiments.} \label{fig2}
\end{figure}

As can be seen from Figure~\ref{fig1}, the new trapping facility
will be inserted into the re-injection channel between the ESR and
the SIS. The new facility will broaden the current research field
and open up several new ones, as indicated by the list in
Figure~\ref{fig2}. The combination of high charge states and very
low kinetic energies makes it possible to (re)trap the ions in
Penning traps, where the ions themselves can be studied with high
precision, or to collide them with gases and surfaces, in order to
study their interaction with neutral matter \cite{BEI05}.

In order to prepare the HITRAP facility itself as well as the
different experimental set-ups to be used at the HITRAP facility
with low-energy highly charged ions, the HITRAP RTD Network was
created, which was funded from 2001 through 2005 by the European
Union. Within this project, there were different teams (nine in
total) from different countries, each team preparing an
experiment, (see Table~\ref{tab1}), or providing theory. The goal
of the HITRAP network was {\it the development of novel
instrumentation for a broad spectrum of physics experiments with
heavy highly charged ions (up to U$^{92+}$) at low energies ($<1$
eV/u)}. Within this project, instrumentation was developed for
high-precision measurements of atomic and nuclear properties, mass
and $g$-factor measurements, and ion-gas and ion-surface
interaction studies.

The construction of the HITRAP facility, which is being carried
out in close collaboration between the GSI Divisions for
Infrastructure, Accelerators and Atomic Physics and the Institute
for Applied Physics at the University of Frankfurt, started in the
beginning of 2005 when appropriate funds were available.
Commissioning of the HITRAP facility is planned for the first half
of 2008 so that the experimental teams can start to perform their
experiments in the second half of 2008.

\begin{table}[!tb]
\caption{Participating teams in the HITRAP EU RTD Network. (RIMS:
Recoil Ion Momentum Spectroscopy, HFS: HyperFine Structure).}
\centering
\begin{tabular}{|l|l|l|}
\hline
{\bf Team} & {\bf Leader} & {\bf Task} \\
\hline
Darmstadt, DE & H.-J. Kluge, & HITRAP facility \\
& O. Kester & \\
& W. Quint & \\
Orsay, FR & J.-P. Grandin & RIMS (H1)\\
Groningen, NL & R. Morgenstern, & ion-surface exp. (H2)\\
& R. Hoekstra & \\
Mainz, DE & G. Werth, & g-factor, mass (H4,H5)\\
& K. Blaum & \\
Krakow, PL & A. Warczak & x-ray spectroscopy (H3)\\
Stockholm, SE & R. Schuch & mass (H5)\\
Heidelberg, DE & J. Ullrich & RIMS (H1)\\
Vienna, AU & J. Burgd\"orfer & theory (H2)\\
London, UK & R.C. Thompson & HFS (H6) \\
\hline
\end{tabular}
\label{tab1}
\end{table}

\section{The HITRAP facility}
\begin{figure}[!tb]
\centering
\includegraphics[width=8.5cm]{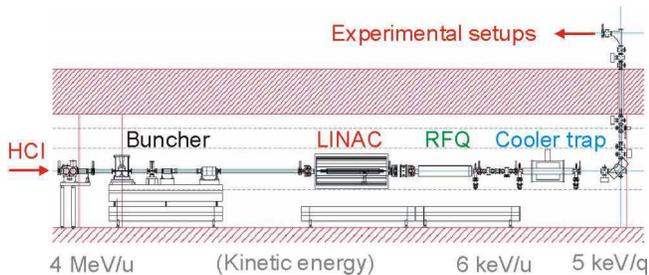}
\caption{Side view of the HITRAP facility. The highly charged ions
come from the left and are decelerated by LINAC and RF quadrupole
structures. They are injected at an energy of 6 keV/u into the
Cooler Penning trap where they are trapped and cooled. After
cooling to liquid-helium temperature they are ejected and
transported with very low energies to the experiments on top of
the platform.} \label{fig3}
\end{figure}

The new trapping facility is shown schematically in
Figure~\ref{fig3}. The highly charged ions are accelerated in the
SIS to typically 400 MeV/u, almost completely stripped and
injected into the ESR. Here they are electron cooled as well as
decelerated and a single bunch is created by bunch-merging. At an
energy of 4 MeV/u, the ions are ejected out of the ESR as a bunch
of about $10^5$ ions, with a pulse length of 1 $\mu$s (roughly 1 m
long) and enter the linear decelerator of HITRAP. The facility is
designed to operate in a pulsed mode, which means that this cycle
(filling the ESR, cooling, deceleration and ejection) will be
repeated every 10 seconds. Before the ion bunch enters the LINAC
and the radiofrequency quadrupole (RFQ) structure, the ion pulse
is reshaped by the buncher (Figure~\ref{fig3}). After deceleration
in the RFQ, the ions enter the Cooler (Penning) trap with only a
few keV/u.

The Cooler trap operates in two steps for cooling the highly
charged ions to liquid-helium temperature: electron cooling and
resistive cooling. For the first step, there are sections with
cold electrons that interact with the highly charged ions, thus
dissipating the ions' kinetic energy. The electrons themselves
cool by synchrotron radiation inside the cold bore of the
superconducting magnet within a few seconds \cite{Bern}. In the
case of resistive cooling, the ions induce image charges in the
trap electrodes. Connecting a frequency-resonant RLC-circuit to
the trap electrodes allows for energy dissipation to this external
cryogenic circuit. Resistive cooling is expected to take several
seconds \cite{Maero}. Finally, the trapped highly charged ions
will have a thermal energy corresponding to slightly more than 4
K, due to noise in the electronic circuits. The cold ions are then
transported, with kinetic energies of only a few keV/q, to the
different setups installed on top of the re-injection channel as a
high-quality ion beam.

\begin{figure}[!tb]
\centering
\includegraphics[width=8.5cm]{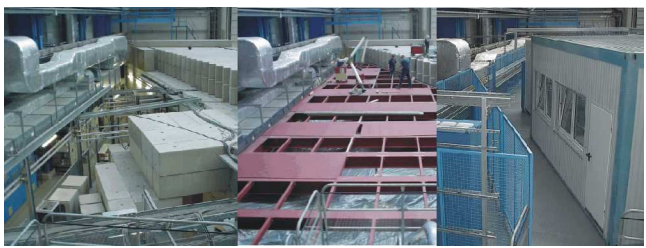}
\caption{Photographs taken from the same position in the ESR
Experimental Hall and showing the construction of the HITRAP
platform. Left: situation in 2004 before construction. Middle:
situation in 2005 when the platform was erected for the huts
housing the electronics of the HITRAP facility and of the the
HITRAP experiments, for RF power supplies and other devices. Right
panel: situation in 2006 when the platform is ready, the huts are
in place, and the installations are almost completed.}
\label{fig4}
\end{figure}

\begin{figure}[!bt]
\centering
\includegraphics[width=8.5cm]{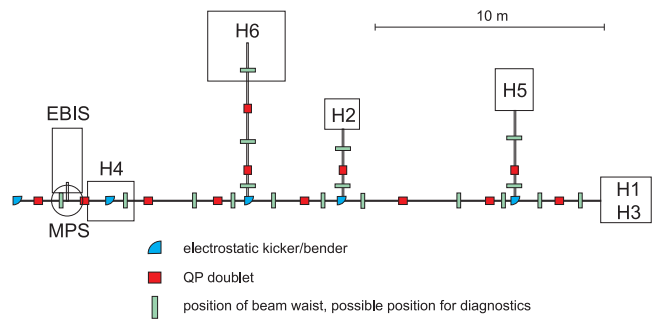}
\caption{Schematic overview of the beamline on the experimental
platform. The experiments are indicated by the numbers H1-H6 (see
Table \ref{tab1}). Beam transport calculations have been performed
leading to the arrangement of ion optical elements as shown in the
figure. The ion beam enters this section from below and travels
from the extreme left hand side to the right. The EBIS is an
off-line ion source for tests of the systems, the MPS
(multi-passage spectrometer) will feed the EBIS ion beam into the
system when needed.} \label{fig5}
\end{figure}

A new platform was built inside the ESR experimental hall at GSI
to house huts for electronics for the local control of the HITRAP
facility and for the different HITRAP experiments and to provide
space for the required infrastructure of the HITRAP facility such
as supplies for radiofrequency, electricity and water. A hole was
drilled through the concrete above the re-injection channel, which
will be used for the beamline to transport the ions from the
Cooler trap towards the setups. The construction of this platform
is illustrated by the photo series in Figure~\ref{fig4}.

For the HITRAP experiments, a complete network of beamlines and
ion optical elements needs to be developed and constructed.
Calculations of these elements and the beam transport have been
carried out showing that nearly 100\% transmission is possible. A
scheme of the current layout of the experiments is shown in
Figure~\ref{fig5}. Most of the ion optical elements are designed.
The elements for bending the ion beam, the so-called
kicker-benders, are under construction. The beamline towards the
experiments needs to have a vacuum of the order of $10^{-11}$ mbar
or better to avoid charge exchange with residual gas. This puts
severe demands on the materials used for the construction of the
ion optical elements almost approaching those encountered when
constructing an electrostatic UHV storage ring such as {\it e.g.}
ELISA in Aarhus, Denmark \cite{MOL97}.

\begin{table}[!th]
\caption{The time schedule for the HITRAP project.}
\centering
\begin{tabular}{|l|l|}
\hline
{\bf Action} & {\bf Date} \\
\hline
Preparation of infrastructure & Ongoing \\
(safety, media supply, controls, RF) & \\
\hline
Delivery of the Cooler trap magnet & Spring 2007 \\
\hline
Test of HITRAP low-energy beam & Mid 2007 \\
transport and Cooler trap with EBIS & \\
\hline
Installation of buncher cavities in the & Spring 2007 \\
re-injection channel & \\
\hline
First test of buncher cavities with beam & May 2007 \\
from ESR & \\
\hline
Installation of LINAC cavities & Fall 2007 \\
\hline
Commisioning of the LINAC & Spring 2008 \\
\hline
Commissioning of HITRAP & Spring 2008 \\
\hline
First experiments & 2008 \\
\hline
\end{tabular}
\label{tab2}
\end{table}

Because the HITRAP facility will not be able to continuously
obtain the ESR beam for testing the Cooler trap and the
experimental setups, it was decided to install an electron beam
ion source (EBIS), which was available from the University of
Frankfurt, on the platform for off-line tests. This source
(MAXEBIS) is currently being tested and will be used to tune the
low-energy beam line (LEBT) towards the Cooler trap, and for tests
of the Cooler trap itself. Afterwards, the components will be
moved to the HITRAP facility. A more complete overview of the time
planning for the construction of the HITRAP facility is presented
in Table~\ref{tab2}.

\section{Experiments at HITRAP}
Since this publication is dedicated to Ingvar Lindgren on the
occasion of his 75$^{th}$ birthday, out of the many experiments
becoming possible (see inset of Figure~\ref{fig2}), we discuss
here only those which are very close to the heart of Ingvar
Lindgren. These are the g-factor of the bound electron, the
hyperfine structure, and the precision mass measurements. X-ray
spectroscopy is discussed in a publication by T. St\"ohlker {\it
et al.} in the same volume \cite{ST07}. In all these experiments,
quantum electrodynamics plays a dominant role, and the progress in
atomic theory is essential.

\begin{figure}[!th]
\centering
\includegraphics[width=6cm]{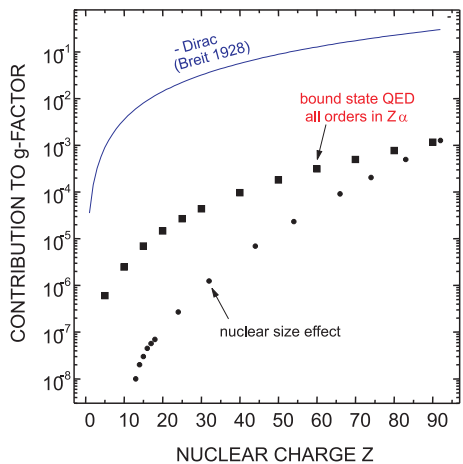}
\caption{Contributions to the $g$-factor versus nuclear charge
$Z$. For heavy ions, the situation is more difficult, but also
more interesting, since the QED and nuclear size contributions
become comparable.} \label{fig6}
\end{figure}

\subsection{The $g$-factor of the bound electron}
The precise measurement of the $g$-factor of the electron bound in
a hydrogen-like ion is a sensitive test of bound-state QED at high
fields. Highly charged ions are ideal for such studies, because
the electromagnetic fields existing close to the nucleus are much
higher than those which can be created artificially in a
laboratory. For example, the electric field strength close to the
nucleus of U$^{91+}$ is of the order of $10^{16}$ V/cm, which is
higher than those created by the strongest lasers available. At
such high electromagnetic fields, perturbative QED calculations
are no longer accurate, and higher-order terms need to be
evaluated carefully. Unfortunately, in this regime there is little
or no experimental data to compare the non-perturbative codes to.
This is precisely why $g$-factor experiments are important.

In relativistic Dirac theory, the $g$-factor of an electron bound
to a H-like ion is given by \cite{BREI28,BEI00}
\begin{equation}
g=\frac{2}{3} \left( 1 + 2 \sqrt{1-(Z \alpha)^2} \right),
\label{eq1}
\end{equation}
where $Z$ is the nuclear charge and $\alpha$ the fine structure
constant. The ratio of the bound-electron ($g_{bound}$) to the
free-electron $g$-factor ($g_{free}$) can be expressed, to leading
order in $Z \alpha$, as
\begin{equation}
\frac{g_{bound}}{g_{free}} \approx 1- \frac{1}{3} (Z \alpha)^2 + \frac{1}{4 \pi} \alpha (Z \alpha)^2.
\label{eq2}
\end{equation}

The first two terms in Eq.~(\ref{eq2}) are dominant and stem from
Dirac theory, the third term comes from bound-state QED. Both
contributions are indicated in Figure~\ref{fig6}, and it can be
seen that at high $Z$ the QED term is of the same order of
magnitude as the nuclear size effects. From Eq.~(\ref{eq1}) it is
clear that for $Z$ larger or equal to $\alpha^{-1}$, $g$ is
undefined. Already when $Z \ll \alpha^{-1}$ is no longer valid,
non-perturbative calculations have to be performed
\cite{BLU97,PER97}. Corresponding first calculations were
initiated by I. Lindgren and G. Soff.

\begin{figure}[!tb]
\centering
\includegraphics[width=7cm]{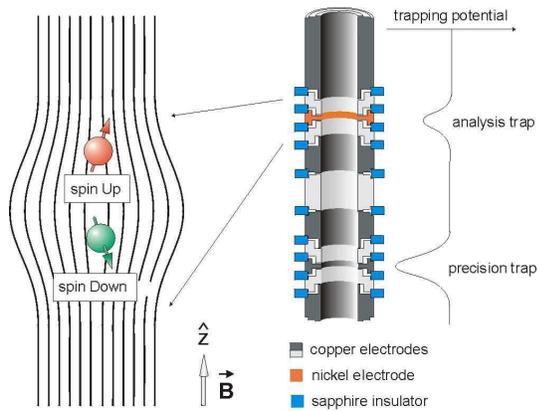}
\caption{Schematic of the double Penning trap setup used for
$g$-factor measurements of the bound electron.} \label{fig7}
\end{figure}

The $g$-factor of the bound electron in $^{12}$C$^{5+}$
\cite{HAF00} and $^{16}$O$^{7+}$ \cite{VER04} has been obtained
via spin-flip measurements of a single cold (4 K) ion. The setup
was constructed in a GSI-Mainz collaboration, and consists of two
Penning traps \cite{HAF03}, {\it i.e.} a `precision trap' and an
`analysis trap', placed in one superconducting magnet. A schematic
of the trap is shown in Figure~\ref{fig7}. The $g$-factor can be
obtained from measurements of the cyclotron frequency
$\omega_c=qB/M_i$ and the Larmor frequency $\omega_L=geB/(2m_e)$,
since it can be expressed as
\begin{equation}
g=2 \left( \frac{q}{e} \right) \left( \frac{m_e}{M_i} \right) \left(\frac{\omega^e_L}{\omega^i_c} \right),
\label{eq3}
\end{equation}
where $m_e,e$ and $M_i,q,$ are the mass and charge of the electron
and the ion, respectively. Using high-quality resonant circuits,
the oscillation frequencies of an ion inside a Penning trap can be
measured independently and with high accuracy. For these
measurements, the polarisation of the electron spin is 100\%,
because only one ion is used. The cyclotron frequency $\omega_c$
is determined inside the precision trap, and a spin-flip may be
induced by microwave irradiation at a frequency close to the
Larmor frequency $\omega_L$. Detection of a spin-flip takes place
after transporting the ion to the analysis trap. Here, a
significant inhomogeneity of the magnetic field, produced by a
nickel ring, makes the $z$-motion of the ion sensitive to the spin
direction. The frequency of the $z$-motion is detected via
electronic detection (based on image charges). After analyzing the
spin direction, the ion is transported back to the precision trap
for the next measurement cycle.

\begin{figure}[!bt]
\centering
\includegraphics[width=8cm]{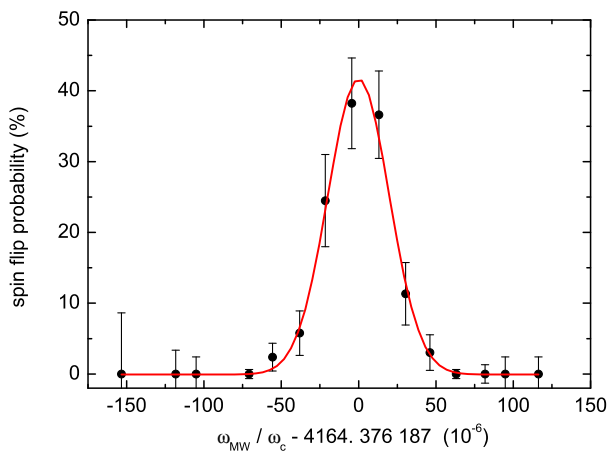}
\caption{Larmor resonance of the $g$-factor measurement of the
bound electron in $^{12}$C$^{5+}$ \cite{HAF00}.} \label{fig8}
\end{figure}

The experimental result obtained for a $g$-factor measurement of a
single $^{12}$C$^{5+}$ ion is presented in Figure~\ref{fig8}. The
horizontal axis shows the ratio of the microwave frequency
$\omega_{MW}$ to the cyclotron frequency $\omega_c$, from which
the $g$-factor of the bound electron can be obtained. The
experimental $g$-factors of the hydrogen-like ions $^{12}$C$^{5+}$
\cite{HAF00} and $^{16}$O$^{7+}$ \cite{VER04} agree within the
uncertainties, which are dominated by the accuracy of the electron
mass, with the calculated values \cite{BEI00,LIND}. The $g$-factor
of the $1s$ electron thus enabled a test of bound-state QED at a
level of 0.25\%.

\begin{figure}[!bt]
\centering
\includegraphics[width=8.5cm]{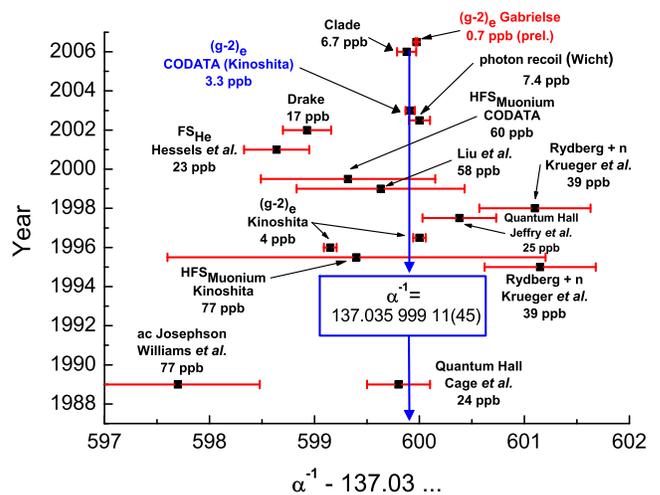}
\caption{Landscape of the fine structure constant $\alpha$: the
evolution of the different measurements and their accuracies
during the last 20 years is indicated.} \label{fig9}
\end{figure}

Accurate $g$-factor measurements also provide means to obtain
better values for fundamental constants, such as the fine
structure constant $\alpha$ (Figure~\ref{fig9}). For example,
because of the good agreement with theory, the $g$-factor data led
to a determination of the electron mass with a four times higher
accuracy \cite{BEI02}. The relative experimental accuracy $\delta
m/m$ of this measurement was as good as $6 \times 10^{-10}$.

From Eq.~(\ref{eq1}) it can be seen that the $g$-factor is linked
to $\alpha$, and it can easily be shown that the relative
uncertainty in $\alpha$ is correlated to that in $g$ via
\begin{equation}
\label{eq4}
\frac{\delta \alpha}{\alpha} \propto \frac{1}{(Z \alpha)^2} \frac{\delta g}{g}.
\end{equation}
It is therefore most interesting to measure $g$ at the highest
possible $Z$. Succeeding the measurements in $^{12}$C$^{5+}$
\cite{HAF00} and $^{16}$O$^{7+}$ \cite{VER04}, a measurement of
the electronic $g$-factor in $^{40}$Ca$^{19+}$ \cite{VOG04} is
currently being prepared. When HITRAP is operational, similar
experiments on heavy systems like $^{238}$U$^{91+}$ will be
performed. It is expected that, from such a measurement, $\alpha$
(Figure~\ref{fig9}) can be obtained with an accuracy of $10^{-8}$
\cite{shabaev02}. A combination of such measurements in H-like and
B-like heavy ions can increase this accuracy significantly
\cite{sha06}.

\subsection{Precision mass measurements}
Masses of stable or radioactive nuclides can be measured with very
high accuracy and single-ion sensitivity, using Penning traps
\cite{LUN03,BLA06}. Since the cyclotron frequency of a trapped ion
increases with the charge state, highly charged ions provide
better resolution and potentially also higher accuracy than singly
charged ions. The SMILETRAP group in Stockholm has pioneered the
use of highly charged ions in ion traps for mass spectrometry
\cite{JER91,FRI06,NAG06}. They measured the masses of several
ions, which are important for tests of QED or double-beta decay,
with an uncertainty close to $10^{-10}$. Using the mass of
$^{12}$C$^{6+}$, Van Dyck {\it et al.} \cite{DYC06} in Seattle
measured the masses of stable singly charged ions with an accuracy
of about $10^{-10}$. A similar accuracy was obtained for the
proton-antiproton comparison at CERN, measured by the group of
Gabrielse \cite{GAB06}. Pritchard {\it et al.} \cite{THO04} at MIT
even had a mass spectrometer, which is now at Florida State
University \cite{RED06}, with an accuracy of about $10^{-11}$. The
masses of singly charged radionuclides have been measured with an
accuracy of $10^{-7}$ to $10^{-8}$ and better, by Penning trap
mass spectrometers installed at ISOLDE/CERN \cite{HER01,MUK},
Argonne \cite{SAV06}, Jyv\"askyl\"a \cite{ERO06}, MSU
\cite{BOL06a}, and GSI \cite{RAU06}. Highly charged ions for mass
spectrometry will be used at TITAN at ISAC/TRIUMF, Vancouver
\cite{DIL06}, LEBIT/MSU \cite{BOL06b}, HITRAP/GSI \cite{HER06a},
ISOLTRAP/ISOLDE \cite{HER06b}, MAFFTRAP/Munich \cite{HAB06}, and
at MATS/GSI \cite{BLA06}.

Table~\ref{tab3} shows a comparison between a mass measurement in
a Penning trap with a singly charged ion, and one with a highly
charged ion. The numbers show that, for the same isotope, an ion
with a charge of $q=50$ already leads to an improved mass
resolution of nearly two orders of magnitude. Longer observation
times (for example, $T_{obs} \geq 10$ s) and higher charge states
({\it i.e.} $q=92$ for uranium) can further improve the mass
resolution. In principle, a mass measurement accuracy of
$10^{-11}$ or better can be reached. This would, for example, make
it possible to `weigh' the $1s$ Lamb shift in U$^{91+}$ with an
accuracy better than presently possible by x-ray spectroscopy
\cite{GUM05}.

\begin{table}[!bh]
\centering \caption{Comparison of the accuracy of mass
measurements for a singly and a highly charged ion with mass
$A=100$ in a magnetic field of $B=6$ T. $\nu_c$ is the cyclotron
frequency, $T_{obs}$ the observation time of a measurement, $R$ is
the resolving power given by the ratio $\nu_c / \delta \nu_c$, and
$\delta m/m$ is the relative mass uncertainty.}
\begin{tabular}{|l|l|}
\hline
\multicolumn{2}{|c|}{singly charged ion} \\
\hline
$q=1$ & $\nu_c=1$ MHz \\
$T_{obs}=1$ s & $\delta \nu_c=1$ Hz \\
$R=10^6$ & $\delta m/m \approx 10^{-8}$ \\
\hline
\multicolumn{2}{|c|}{highly charged ion} \\
\hline
$q=50$ & $\nu_c=50$ MHz \\
$T_{obs}=1$ s & $\delta \nu_c=1$ Hz \\
$R=5 \times 10^7$ & $\delta m/m \approx 2 \times 10^{-10}$ \\
\hline
\end{tabular}
\label{tab3}
\end{table}

\subsection{Laser spectroscopy of hyperfine structure}
\begin{figure}[!tb]
\centering
\includegraphics[width=7cm]{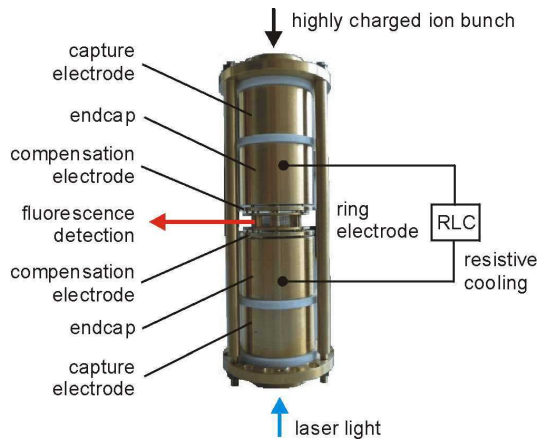}
\caption{Photograph of the spectroscopy trap, which will be used
for laser spectroscopy measurements of hyperfine splittings in
highly charged ions. The ring electrode is split into four
segments, which are covered by highly transparent mesh.}
\label{fig10}
\end{figure}

An experiment is being prepared to perform laser spectroscopy of
ground state hyperfine structure in highly charged ions
\cite{VOG05}. Normally, the hyperfine splitting wavelength in
atoms and ions is in the microwave region. However, since the
hyperfine splitting of hydrogen-like ions scales with the nuclear
charge as $Z^3$, the hyperfine structure becomes measurable by
laser spectroscopy above $Z \approx 60$. This allows for accurate
laser spectroscopy measurements of these hyperfine splittings and,
in turn, for sensitive tests of corresponding calculations of
transition energies and lifetimes. Previous measurements were
carried out for $^{209}\mbox{Bi}^{82+}$ \cite{KLA94},
$^{165}\mbox{Ho}^{66+}$ \cite{CRE96}, $^{185,187}\mbox{Re}^{74+}$
\cite{CRE98}, $^{207}\mbox{Pb}^{81+}$ \cite{SEE98},
$^{209}\mbox{Bi}^{80+}$ \cite{BEI98} and
$^{203,205}\mbox{Tl}^{80+}$ \cite{BEI01}. These measurements
suffered from the Doppler width and shift of the transition, which
were due to the relativistic velocities and velocity spreads of
the ions in the ESR. In the case of the measurements conducted in
the EBIT (electron beam ion trap) at Livermore, resonance
transitions were broadened due to the high temperatures of the
ions in the EBIT. The new experiments will be performed in a
Penning trap with confined and cold highly charged ions at
cryogenic temperatures. This strongly reduces the Doppler effects
and will lead to a measurement resolution of the order of
$10^{-7}$. Hence, these results will allow the determination of
hyperfine anomalies which requires high precision. The trapped
ions will be laser-excited along the trap axis, and fluorescence
detection will take place perpendicular to the trap axis, through
the transparent ring electrode (see Figure~\ref{fig10}).

Furthermore, by use of a rotating wall technique \cite{ITA98},
which radially compresses the ion cloud via a rotating dipole
field applied to the segmented ring electrode, a high ion number
density can be obtained. Together with the localisation, this
enhances the intensity of the measured fluorescence. For an ion
cloud of $10^5$ highly charged ions at the space charge limit of
the trap, we calculated fluorescence rates of several thousand
counts per second on a background of a few hundred counts per
second (S/N $\approx$ 50) \cite{WIN05}. Laser intensities of
several tens of W/m$^2$ are required to saturate the 10 mm$^2$ ion
cloud, which implies moderate laser powers of just a few mW
\cite{VOG05}.

Furthermore, measurements of hyperfine structure in H- and Li-like
ions can test quantum electrodynamics (QED) at high
electromagnetic fields \cite{SHA01}. Such accurate measurements
can be compared to corresponding calculations that include nuclear
effects, such as the Bohr-Weisskopf and Breit-Schawlow effects, as
well as (higher-order) QED effects. From a comparison of the
hyperfine splitting in H- and Li-like ions of the same isotope,
the nuclear effects cancel to a large extent, thus enabling a
stringent test of the QED effects \cite{SHA01}. The necessary
experimental resolution for such a comparison is of the order of
$10^{-6}$, which can relatively easily be met by laser
spectroscopy \cite{WIN06}.

\section{The new GSI facility}
\begin{figure}[h!bt]
\centering
\includegraphics[width=9cm]{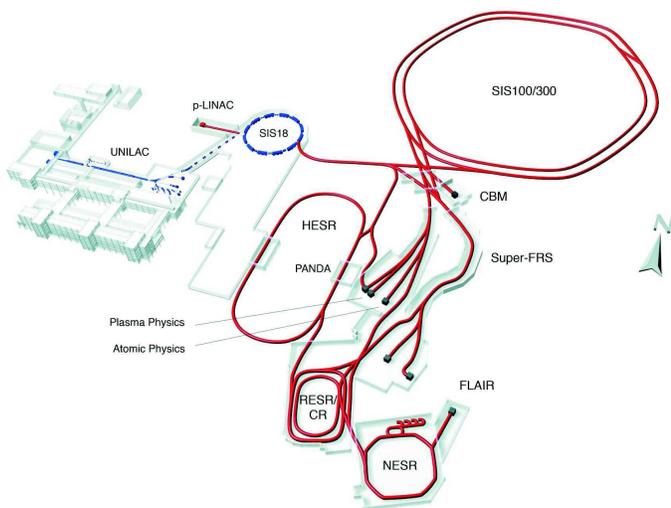}
\caption{The future GSI facility FAIR (Facility for Antiproton and
Ion Research) in Darmstadt.} \label{fig11}
\end{figure}

In the more distant future, HITRAP will be a component of the new
international Facility for Antiproton and Ion Research (FAIR)
\cite{FAI06}. HITRAP will thus, in addition to highly charged and
radioactive ions, also produce low-energy antiprotons.
Furthermore, the FAIR facility will provide the highest
intensities of both stable and radioactive ion beams with energies
up to 34 GeV per nucleon. At such energies, the highly charged
ions generate electric and magnetic fields of exceptional strength
and ultra-short duration. The general layout of FAIR is shown in
Figure~\ref{fig11}.

\subsection{Stored Particle Atomic Research Collaboration (SPARC)}
The new FAIR facility has key features that offer a range of new
opportunities in atomic physics research and related fields, which
will be exploited by the Stored Particle Atomic Research
Collaboration SPARC \cite{STO05}. In particular, the
Superconducting Fragment Separator (SFRS) will provide a rich
spectrum of radionuclides. The high intensity of secondary beams
produced at the SFRS will make it possible to extract decelerated
radioactive ion beams from the New Experimental Storage Ring
(NESR) and to decelerate them for trap experiments with sufficient
intensity at HITRAP. Therefore, the physics program of HITRAP can
be extended to novel experiments with trapped radioactive ions
and, of course, with trapped antiprotons. Trapped radioactive ions
in high charge states may reveal a completely new domain for
fundamental interaction studies and for experiments on the
borderline between atomic and nuclear physics.

Moreover, the manipulation of trapped radioactive ions with laser
light opens up possibilities to study questions of the Standard
Model. By optical pumping within the hyperfine levels of the
ground state, the nuclear spins of radioactive nuclides can be
polarized with high efficiency. The detection of the asymmetry of
beta decay, for example, will allow one to explore deviations from
the vector/axial-vector (VA)-structure of the weak interaction and
to set limits for the masses of heavy bosons, which are not
included in the Standard Model.

Direct mass measurements on unstable nuclides with ultra-high
accuracy (up to $\delta m/m \approx 10^{-11}$) will also be
possible. Such an accuracy would allow one to determine the
binding energy of U$^{91+}$ with an accuracy of $\delta mc^2
\approx 2 eV$. If the QED calculations are found to be correct,
nuclear charge radii of unstable nuclides can be determined.

For a general exploration of masses in the chart of nuclei, a mass
resolution of $\delta m/m \approx 10^{-6}$ to $10^{-7}$ is
sufficient. This is also planned by isochronous or Schottky mass
spectrometry experiments at the New Experimental Storage Ring
(NESR). However, in some cases, like double-beta decay or tests of
the unitarity of the Cabibbo-Kobayashi-Maskawa matrix, a much
higher accuracy is required, which is possible by using trapped
highly charged ions at HITRAP \cite{HER06a} and at MATS
\cite{BLA06}.

\subsection{Facility for Low-Energy Antiproton and Ion Research (FLAIR)}
The planned FLAIR facility will be the most intense source of
low-energy antiprotons world-wide \cite{WID05}. The beam intensity
of extracted low-energy antiprotons will be two orders of
magnitude higher than that of the Antiproton Decelerator (AD) at
CERN. We therefore anticipate that experiments with trapped
antiprotons, which are currently impossible anywhere else (due to
insufficient intensities), will be performed at the FLAIR
facility. A possible highlight in the field of low-energy
antimatter research would be the first direct experimental
investigation of the gravitational interaction of antimatter,
which has never been attempted up to now. Such investigations
could be performed on ultra-cold antihydrogen atoms \cite{WAL04}
which are produced by recombining trapped antiprotons with
positrons in a so-called nested Penning trap. The effect of
gravity on antimatter is an important issue for the development of
quantum theories of gravity.

\acknowledgments We acknowledge support by the European Union, the
German Ministry for Education and Research (BMBF) and the
Helmholtz Association (HGF).

\end{document}